\documentclass[sn-mathphys]{sn-jnl}% Math and Physical Sciences Reference Style
\jyear{2021}%

\theoremstyle{thmstyleone}%
\theoremstyle{thmstyletwo}%

\theoremstyle{thmstylethree}%

\usepackage{csquotes}
\usepackage{hyperref}
\usepackage{bm}
\usepackage{comment}

\usepackage[normalem]{ulem}

\begin{document}

\title[Heisenberg's uncertainty principle and particle trajectories]{Heisenberg's uncertainty principle and particle trajectories}

%%=============================================================%%
%% Prefix	-> \pfx{Dr}
%% GivenName	-> \fnm{Joergen W.}
%% Particle	-> \spfx{van der} -> surname prefix
%% FamilyName	-> \sur{Ploeg}
%% Suffix	-> \sfx{IV}
%% NatureName	-> \tanm{Poet Laureate} -> Title after name
%% Degrees	-> \dgr{MSc, PhD}
%% \author*[1,2]{\pfx{Dr} \fnm{Joergen W.} \spfx{van der} \sur{Ploeg} \sfx{IV} \tanm{Poet Laureate} 
%%                 \dgr{MSc, PhD}}\email{iauthor@gmail.com}
%%=============================================================%%

\author*[1]{\fnm{Serj} \sur{Aristarhov}}\email{S.Aristarkhov@campus.lmu.de}

%\author[2,3]{\fnm{Paula} \sur{Reichert-Sch\"urmer}}\email{paula.reichert@gmail.com}
%\equalcont{These authors contributed equally to this work.}

\affil[1]{\orgdiv{Mathematical Institute}, \orgname{Ludwig Maximillian University Munich}, \orgaddress{\street{Theresienstr. 39}, \city{Munich}, \postcode{80333}, \state{Bavaria}, \country{Germany}}}

%\affil[2]{\orgdiv{Mathematical Institute}, \orgname{Ludwig Maximillian University Munich}, \orgaddress{\street{Theresienstr. 39}, \city{Munich}, \postcode{80333}, \state{Bavaria}, \country{Germany}}}

%%==================================%%
%% sample for unstructured abstract %%
%%==================================%%

\abstract{In this paper we critically analyse W.\ Heisenberg's arguments against the ontology of point particles following trajectories in quantum theory, presented in his famous 1927 paper and in his Chicago lectures (1929). Along the
way, we will clarify the meaning of Heisenberg’s uncertainty relation and
help resolve some confusions related to it.}


%%\pacs[JEL Classification]{D8, H51}

%%\pacs[MSC Classification]{35A01, 65L10, 65L12, 65L20, 65L70}

\maketitle

\section{Introduction}\label{sec1}

Heisenberg's work on quantum mechanics, in particular, his well-known 1927 paper~\cite{heisenberg_1927} and his 1929 Chicago lectures~\cite{chicago}, led to the rejection of the ontology of point particles moving on trajectories in quantum theory.
%The Heisenberg uncertainty relation (HUR) is often said to be incompatible with an ontology of point particles following trajectories in quantum theory. 
As M.\ Born put it in his 1954 Nobel lecture~\cite{born_nobel}:

\begin{quote}
\textit{It was through this paper [Heisenberg's 1927~\cite{heisenberg_1927}] that the revolutionary character of the new conception
became clear. It showed that not only the determinism of classical physics
must be abandoned, but also the naive concept of reality which looked upon
the particles of atomic physics as if they were very small grains of sand. At
every instant a grain of sand has a definite position and velocity. This is not
the case with an electron.}
\end{quote}
%If its position is determined with increasing accuracy, the possibility of ascertaining the velocity becomes less and vice versa

Ever since, this view has become deeply entrenched in many textbooks and lectures (see, e.g., \cite{landau}) and remains popular today.
%% Surprisingly, I could not find more references, although there has to be a ton.

At the same time, the existence and successes of trajectory-containing quantum theories (TCQTs), for instance, Bohmian mechanics (also known as the de Broglie--Bohm or pilot-wave theory)~\cite{red_book, bohm_hiley, holland}, tell us that there must be something wrong with the reasoning of Heisenberg and other advocates of the view that there are no point particles following trajectories in quantum physics.  
%the HUP and particle trajectories in the quantum realm are incompatible. 

In this paper, we will investigate Heisenberg's arguments against the ontology of point particles in quantum mechanics, presented in his well-known 1927 paper~\cite{heisenberg_1927} and his 1929 Chicago lectures~\cite{chicago}. In what follows, we will also clarify the meaning of Heisenberg's uncertainty relation (UR) and help resolve some confusions related to it.

\section{Heisenberg's arguments in 1927}
\label{1927}
It is often asserted~\cite[Ch. 10]{debate}, \cite[p. 100]{bohm} that the impossibility of trajectories in the quantum world is a consequence of the Heisenberg UR (eq.\ \eqref{kennard}, Sec.\ \ref{ur}). Indeed, a relation similar in form to~\eqref{kennard} appears for the first time in \cite{heisenberg_1927}. However, it seems Heisenberg's argument against trajectories \emph{did not} rely on it.

In~\cite{heisenberg_1927} Heisenberg proposed to redefine the familiar physical notions of position, velocity, and trajectory:
%Heisenberg offered to ban the `unmeasurable' entities from quantum theory:
\begin{quote}
\textit{When one wants to be clear about what is to be understood by the words `position of the object,' for example of the electron $<$...$>$, then one must specify definite experiments with whose help one plans to measure the `position of the electron'; otherwise this word has no meaning. }
\end{quote}
According to his new definitions, these words are mere place holders for certain laboratory operations. In the event that no suitable laboratory operations could be found, the corresponding notion was declared meaningless, and had to be banned from the very formulation of the theory. 

%Heisenberg's argument against particle trajectories of 1927 is based on this philosophical principle, and does not use the uncertainty relation (UR), which appeared for the first time in the same paper. 
Regarding the possibility of measuring the trajectories of the electrons, Heisenberg wrote:
\begin{quote}
\textit{By path we understand a series of points in space (in a given reference system) which the electron takes as `positions' one after the other. As we already know what is to be understood by `position at a definite time,' no new difficulties occur here. Nevertheless, it is easy to recognize that, for example, the often used expression, the `1s orbit of the electron in the hydrogen atom,' from our point of view has no sense. In order to measure this 1s `path' we have to illuminate the atom with light whose wavelength is considerably shorter than $10^{-8}$ cm. However, a single photon of such light is enough to eject the electron completely from its `path' $<$...$>$.}
\end{quote}
Thus he admits that it \emph{is} possible to measure particle trajectories, at least in certain situations.\footnote{For instance, in the debate with Einstein~\cite[Ch. 10]{debate}, Heisenberg concedes that the trajectory of a free particle can be measured (say, approximately) via the Wilson cloud chamber.} It would require a series of consecutive position measurements and \emph{not} the simultaneous determination of the position and momentum of the particle. So the UR is not what makes the measurement of the `1s orbit' problematic, but rather the assertion that a single measurement of the electron's position via photons would ionize the atom. Therefore, from Heisenberg's point of view, the notion of an electron's trajectory in an atom makes no sense. 

The operational principle used by Heisenberg, which we could call `measurement = meaning' following~\cite{plato}, was criticized for being inconsistent or circular~\cite[p. 73]{jammer}: In order to show that a certain entity (e.g., a particle trajectory) cannot be observed in an experiment, one has to first specify what one means by the said entity. And Heisenberg does exactly that for the trajectory (or the `path') of an electron at the beginning of the above quote. It is inconsistent with the claim that this word (in the narrower context of the hydrogen atom) has no meaning and has to be removed from the theory.  

%\begin{displayquote}
%\textit{At the instant when position is determined -- therefore, at the moment when the photon is scattered by the electron -- the electron undergoes a discontinuous change in momentum. This change is the greater the smaller the wavelength of the light employed -- that is, the more exact the determination of the position. At the instant at which the position of the electron is known, its momentum therefore can be known up to magnitudes which correspond to that discontinuous change. Thus, the more precisely the position is determined, the less precisely the momentum is known, and conversely.}
%\end{displayquote}   

\section{Heisenberg's arguments in 1929}
Heisenberg's Chicago lectures~\cite{chicago} contain an entire chapter devoted to the `critique of the corpuscular theory of matter' (Ch.\ II). As we shall see, the argument against particle trajectories presented therein is significantly different from that of the 1927 paper.

\subsection{The uncertainty relation}
\label{ur}
Chapter II starts with a discussion of
%position-momentum UR in the form we know it today from the popular textbooks (see e.g.~\cite{griffiths}):\footnote{The only difference is that $\Delta a, a=p,q$ in~\cite{chicago} is equal $\sqrt{2}$ times standard deviation and not the standard deviation itself. That is why in~\cite{chicago} the right-hand side of the inequality is $\hbar$ and not the usual $\hbar/2$.}
the inequality
\begin{equation}
\label{kennard}
\Delta q\Delta p\geq \hbar,
\end{equation}
where $\smash{(\Delta q)^2=\int (q'-\bar{q})^2\vert\psi_q(q')\vert^2\mathrm{d}q'}$ with $\smash{\bar{q}=\int q'\vert\psi_q(q')\vert^2\mathrm{d}q'}$, and analogously for $\Delta p$, where $\psi_q$ and $\psi_p$ are the wave function of the particle and its Fourier transform, respectively.\footnote{Just like Heisenberg, we assume that the particle moves in one-dimensional space for brevity. The generalization to three-dimensional motion is straightforward.} 
This inequality is a mathematical fact, which was first proven, notably, not by Heisenberg, but by Kennard~\cite{kennard} in 1927. 
%Kennard's result was later generalized by Robertson~\cite{robertson} and Schr\"odinger~\cite{schroedinger}.
In modern textbooks (see, e.g.,~\cite{griffiths}), it is called the Heisenberg position--momentum UR.\footnote{The only difference is that $\Delta a, a=p,q$ in~\cite{chicago} is equal to $\sqrt{2}$ times the standard deviation and not the standard deviation itself. That is why in~\cite{chicago} the right-hand side of the inequality is $\hbar$ and not the usual $\hbar/2$.} 

%The empirical import of this inequality is clear: $\Delta q$ and $Delta p$ approximate the standard deviations of the positions and momenta of a particle measured in two different sufficiently large ensembles of identically prepared single-particle systems.   take an ensemble of identically prepared single-particle-system, prepare it in a certain way, let it evolve a definite amount of time and measure the position $q$ of the particle. Then we repeat the same experiment large number of times, taking particles identical to the first one and preparing them similarly. As the result we obtain the distribution of the results, whose mean squared error is known to be close to $(\Delta q)^2$.

The empirical import of this inequality is clear: Take two large ensembles of identically prepared single-particle systems. For the first ensemble, measure the particle position in each system. Then $\Delta q$ approximates the standard deviation of the measured positions. For the second ensemble measure the momentum of the particle in each system. The resulting standard deviation then approaches $\Delta p$. The inequality~\eqref{kennard} thus states that the product of those two standard deviations cannot be smaller than $\hbar$. 

%a single-particle-system, prepare it in a certain way, let it evolve a definite amount of time and measure the position $q$ of the particle. Then we repeat the same experiment large number of times, taking particles identical to the first one and preparing them similarly. As the result we obtain the distribution of the results, whose mean squared error is known to be close to $(\Delta q)^2$. Then we repeat this whole process again, but this time we perform the `momentum measurements' -- measurements, whose statistics by definition is dictated by the Fourier transform of the wave function of the particle $\psi_q$ at the time of the measurement. The mean squared error of the distribution of the results will be close to $(\Delta p)^2$.
%Therefore the inequality~\eqref{kennard} tells us that the product of the widths of those two empirical distributions cannot be smaller than $\hbar$.

On page 15 of the Chicago lectures, Heisenberg writes:
\begin{quote}
\textit{This UR [\eqref{kennard}] specifies the limits within which the particle picture may be applied. Any use of the words `position' and `velocity' with an accuracy exceeding that given by equation (I) [\eqref{kennard}] is just as meaningless as the use of words whose sense is not defined.} 
\end{quote}

These words and the preceding heuristic derivation of~\eqref{kennard} probably gave birth to the following argument against the ontology of point particles~\cite[Ch. 10]{debate},~\cite{bohm}.\footnote{This argument seems to be particularly widespread among laypersons and in the popular science media.} Since quantum particles are known to be wave-like, we may ascribe a definite position to a particle only if the corresponding wave $\psi(x)$\footnote{ A one-dimensional picture is assumed for simplicity.} is sharply peaked. At the same time, it follows from~\eqref{kennard} that, the narrower $\psi(x)$, the broader its Fourier transform $F[\psi](k)$. But according to the de Broglie relation, we have $p=k\hbar$, so a matter wave which is well localized in space, thus resembling a particle with a definite position, cannot have a definite momentum, and vice versa.

As noted by Ballentine~\cite{ballentein}, this conclusion rests on the identification of the particle with the wave packet itself, which, he argues, is unjustified given the example of a particle incident on a beam splitter with detectors placed on either side. The particle is either reflected or transmitted, whereas the wave packet is separated into reflected and transmitted components.

Here, it is worth emphasizing that, even if it were somehow possible to formulate a version of quantum theory based on the identification of particles with wave packets, which would reject the notion of particle trajectories by the given reasoning, it is not at all certain that such a version would be in any way advantageous in comparison to TCQTs like Bohmian mechanics.

\subsection{Heisenberg's argument}
\label{argument}
Although the Heisenberg quote from the previous subsection could be interpreted in the manner described above, it seems that Heisenberg's own argument against particle trajectories was markedly different. On page 20, he states:

\begin{quote}
\textit{$<$...$>$ if the velocity of the electron is at first known and the position then exactly measured, the position for times previous to the measurement may be calculated. Then for these past times $\Delta p\Delta x$ is smaller than the usual limiting value, but this knowledge of the past is of a purely speculative character, since it can never (because of the unknown change in momentum caused by the position measurement) be used as an initial condition in any calculation of the future progress of the electron and thus cannot be subject to experimental verification. It is a matter of personal belief whether such a calculation concerning the past history of the electron can be ascribed any physical reality or not.}
\end{quote}

Popper~\cite[p. 27]{popper} criticized Heisenberg's conclusion here, pointing out that most measurements, especially in quantum physics, are retrodictive. It is a standard praxis to measure, say, the position of a particle to determine some of its past properties (energy or momentum) using the theory and knowledge of the experimental arrangement. Thus, according to Popper, `To question whether the so ascertained ``past history of the electron can be ascribed any physical reality or not'' is to question the significance of an indispensable standard method of measurement'.

Popper was certainly right, but it seems that he missed Heisenberg's point, especially if we take into account the footnote on page 15:\footnote{It is worth mentioning that this footnote comes right after the words `This uncertainty relation...', quoted in subsection~\ref{ur}, which confirms that the argument against trajectories that Heisenberg had in mind in the Chicago lectures was not the one based on the identification of the electron with a wave.}
\begin{quote}
\textit{In this connection one should particularly remember that the human language permits the construction of sentences which do not involve any consequences and which therefore have no content at all -- in spite of the fact that these sentences produce some kind of picture in our imagination; e.g., the statement that besides our world there exists another world, with which any connection is impossible in principle, does not lead to any experimental consequence, but does produce a kind of picture in the mind. Obviously such a statement can neither be proved nor disproved.}
\end{quote}
It becomes clear that Heisenberg did not question the possibility of reconstructing the trajectories once we assume their existence.\footnote{This has in fact been done for photons~\cite{steinberg}. A similar procedure may be possible for massive particles~\cite{vel_mes}.} He claimed that as a consequence of the UR we cannot `verify' their existence. Hence, he claims, `it is a matter of personal belief' to accept or to reject the results of retrodictive measurements performed under the assumption of their existence. 

By `experimental verification' Heisenberg apparently meant the following: At time $t=0$ we determine (set) the position of the particle to be $q_0$ with an accuracy $\epsilon_q$ and determine other necessary initial data (for Heisenberg it is the momentum) to predict the future position of the particle. Let $U_0$ be the region of space where the probability of finding the particle is close to one. This region can be thought of as a ball of radius $\epsilon_q$ around $q_0$. 
%We find the image $U_\tau$ of the ball $B_{\epsilon_q}(q_0)$ of the radius $\epsilon_q$ around $q_0$ under the flow corresponding to the alleged law of motion for the moment of time $\tau>0$.
Let $U_\tau$ be the region of space where the trajectories determined by the alleged law of motion can end up at time $\tau>0$, if they start in the region $U_0$ and the (in)accuracy of other initial data is taken into account. 
The law of motion and the accuracy of the initial data have to be such that the volume of $U_\tau$ is still of order $\epsilon_q$. At time $\tau$ we measure the position of the particle $q(\tau)$ with an accuracy $\epsilon'_q$. Let $B_{\epsilon'_q}(q(\tau))$ be the ball of radius $\epsilon'_q$ with center at $q(\tau)$. If, no matter how small we manage to make $\epsilon_q$ and $\epsilon'_q$, we have $U_\tau \cap B_{\epsilon'_q}(q(\tau))\neq \varnothing$, the `verification' is complete. According to Heisenberg, because of the UR it is impossible to determine the initial data (the momentum) in such a way that $U_\tau$ is of the size $\epsilon_q$, no matter how small the latter is, so the `verification' is impossible. 

It seems to us that the word `verification' is not really appropriate here. In fact, if the volume of $U_\tau$ is large and we have $U_\tau \cap B_{\epsilon'_q}(q(\tau))\neq \varnothing$, we can still claim to have verified our trajectory-containing theory. Indeed, we fix the initial conditions in a certain way (prepare the system), then let it evolve for some time, and our observations match the predictions. Presumably, what Heisenberg meant was the verification of the \textit{necessity} of trajectories in the theory. He thought that, if his `verification' could be carried out, the existence of trajectories would be the only possible explanation of the observation. However, if $U_\tau$ can never be made small, at least one more possibility arises: To leave only the position distribution ($\vert\psi\vert^2$) in the description and banish the trajectories.\footnote{Strictly speaking, this is incorrect. Of course, in theory~\cite{nino} it is possible to set the initial conditions for a classical particle exactly (with zero inaccuracy), but in practice we always have a distribution in phase space. This distribution may be propagated according to the Liouville equation. Thus the trajectory-free description is not excluded even if Heisenberg's `verification' is possible. It is just that, for the center of mass of a large rigid body, the position support of the initially prepared distribution is much narrower than the size of the body. And, no matter how narrow the initially prepared distribution is, it is going to stay narrow with time. So the trajectory-based description feels more natural.} According to Heisenberg it is `a matter of personal belief' which possibility to opt for.
The preferences of Born, Heisenberg, and other members of the Copenhagen school were clearly not on the side of trajectories. Unfortunately, in the following years they often passed their personal tastes off as the absolute truth. In 1958, instead of mentioning that having trajectories in a quantum theory was a real possibility, Heisenberg said something as radical as~\cite[p. 129]{radical_quote}:
\begin{quote}
\textit{The idea of an objective real world whose smallest parts exist objectively in the same sense as stones or trees exist, independently of whether or not we observe them $<$...$>$, is impossible $<$...$>$.}
\end{quote}  
%% Some quotes of other guys to support that?

\subsection{The role of the uncertainty relation}
\label{role}

Let us summarize Heisenberg's argument against trajectories in the Chicago lectures: The UR~\eqref{kennard} makes it impossible to `verify' the existence of the particle trajectories. Therefore it is not necessary to assume it. It is a matter of taste. In the previous subsection we discussed the second part of this argument. Now we are going to focus on the first part and answer the question of how, on the basis of the UR~\eqref{kennard}, Heisenberg concluded that the aforementioned `verification' was impossible. We can then ask whether his conclusion was correct. 

In order to prove this claim, it is necessary to assume the existence of the particle trajectories in the first place. Heisenberg does that implicitly. The law of motion Heisenberg implies has a lot in common with Newton's second law.\footnote{Heisenberg may indeed have intended the law of motion of classical physics. See, for instance the following quote~\cite[p. 65]{heisenberg_1927}: 
`\textit{There must therefore exist for a definite state -- for example, the 1s state -- of the atom a probability function for the location of the electron which corresponds to the mean value for the classical orbit, averaged over all phases $<$...$>$.}'} The initial data determining the trajectory are position and momentum (velocity), so the law has to be second order in time. Apart from that, Heisenberg assumes that the trajectories are straight lines for a free particle. This is clear, for instance, from the first quote given in subsection~\ref{argument}. He also assumes that the distribution of the momentum, meaning the mass times the vector tangent to the trajectory at a given moment of time, of a particle with wave function $\psi$ is given by the Fourier transform of $\psi$.

For Heisenberg, the wave function clearly has an epistemological character. For instance, he writes~\cite[p. 16]{chicago}: 
%It expresses \textit{our knowledge} about the particle.
\begin{quote}
\textit{Any knowledge of the coordinate $q$ of the electron can be expressed by a probability amplitude $S(q')$, $\vert S(q')\vert^2\mathrm{d}q'$ being the probability of finding the numerical value of the coordinate of the electron between $q'$ and $q'+\mathrm{d}q'$.}
\end{quote}
Since the wave function expresses \textit{our knowledge} about the particle, it is obvious for Heisenberg (it follows from the mere definition of the wave function) that, if the particle's position $q$ is known to be $q_0$ within a certain accuracy $\epsilon_q$, its wave function will have `width' $\Delta q\equiv\epsilon_q$. In other words, the wave function of a particle right after a position measurement of accuracy $\epsilon_q$ will have `width' $\Delta q=\epsilon_q$. 

How accurately can the momentum of the particle be determined at the same moment of time (from Heisenberg's point of view)? Suppose it is possible to measure it infinitely accurately before we determine the position. Then the inaccuracy in our knowledge of the momentum is equal to the inaccuracy in the measurement of its disturbance $\delta p$ due to the position measurement.\footnote{Once again, Heisenberg assumes that the momentum, i.e., the mass times the vector tangent to the trajectory, does not change, when the particle is free. In particular, it does not change between the measurements.} Let us call this quantity $\epsilon_{\delta p}$. But, according to Heisenberg's view of the wave function, our knowledge of the momentum of the particle is expressed by the Fourier transform of $\psi$. Its `width' is $\Delta p$. Thus $\epsilon_{\delta p}\equiv\Delta p$ and $\epsilon_q\epsilon_{\delta p}=\Delta q\Delta p\geq \hbar$~\cite[p. 180]{ur_deriv_1927}. Now we recall that the momentum in Heisenberg's considerations is nothing but the vector tangent to the trajectory of the particle and that, if left alone, the particle will move in a straight line in the direction dictated by the initial momentum. Since the indeterminacy in the momentum (our lack of knowledge of the momentum) gets larger as $\epsilon_q$ gets smaller, the region $U_\tau$ -- the set of points the trajectories may end up in at time $\tau$ if they start from the ball $B_{\epsilon_q}(q_0)$,  for the given possible values of the momentum, -- will be large. This was how Heisenberg concluded that his `experimental verification' of the existence of the trajectories was impossible. 

Note that, as already mentioned, the equalities $\epsilon_q=\Delta q$ and $\epsilon_{\delta p}=\Delta p$ are not even assumptions for Heisenberg; due to the assumed epistemological character of the wave function, $\epsilon_q$, $\Delta q$ and $\epsilon_{\delta p}$, $\Delta p$ are equivalent quantities in his view of things. This is why Heisenberg refers to the famous $\gamma$-microscope thought experiment, and other experiments in which both the position and momentum of a particle are determined at the same time (see~\cite{chicago} \textsection 2), as `illustrations' of the UR, even though the experimental situation relating to~\eqref{kennard}, %according to the modern views on the wave function, 
the one we described in subsection~\ref{ur}, does not involve simultaneous measurements and is very different from the examples given by Heisenberg in~\cite{chicago}. For instance, he claimed that, in the case of the $\gamma$-microscope, the position $q$ of the electron can be measured with accuracy equal to the resolving power of the microscope: $\epsilon_q={\lambda}/{\sin\varepsilon}$, where $\varepsilon$ is the angular aperture of the objective. At the same time the change in momentum of the electron due to this measurement (the Compton recoil) $\delta p$ can be found using the theory of the Compton effect and the momentum of the scattered photon. The latter can be measured in the same run of the experiment with accuracy $\epsilon_{\delta p}\sim \frac{h}{\lambda}\sin\varepsilon$. Thus we have $\epsilon_q\epsilon_{\delta p}\sim h$. The same relation is obtained in all other examples of the simultaneous measurement of $q$ and $\delta p$, although they are based on completely different ideas. This was no surprise for Heisenberg because for him it was bound to happen, since $\epsilon_q\equiv \Delta q$, $\epsilon_{\delta p}\equiv \Delta p$ and~\eqref{kennard} holds.
%In all these examples both $q$ and $\delta p$ are determined in the same run of the experiment and in each example one obtains different expressions for $\epsilon_q$ and $\epsilon_{\delta p}$. Nevertheless every time it turns out that  

Now, when we break down Heisenberg's line of reasoning, our aim is to analyse it critically. In order to reach Heisenberg's conclusion, namely the impossibility of `experimentally verifying' the existence of trajectories of quantum particles, it is necessary to consider a generic empirically adequate TCQT. Heisenberg's implicit trajectory-containing quantum theory is not only very special (not generic), but also empirically inadequate. Indeed, Heisenberg postulates that the probability density of the velocity of a particle with the wave function $\psi$ is given by the Fourier transform of $\psi$. Such a velocity field would in general fail to satisfy the continuity equation. Thus the corresponding theory will not reproduce the Born rule. Apart from that, in any TCQT which reproduces the Born rule, the wave function has to be related to the law of motion and will thus have nomological rather than epistemological status. So, if the analysis requires the relations $\epsilon_q=\Delta q$ or $\epsilon_{\delta p}=\Delta p$, which Heisenberg considered to be obvious, they will have to be proven.
%% Example: Gaussian with a complex exponent. Bohmian velocity satisfies the CE and is simply constant at t=0. It is definitely not distributed according to the Fourier transform of this w.f.
Thus Heisenberg's analysis may be correct in itself, but it is irrelevant, as the TCQT he considered did not describe our world.

Although we do not know of any study in which the possibility of Heisenberg's `verification' of the existence of particle trajectories has been checked for the whole class of empirically adequate TCQTs, the relevant analysis for one of those theories, Bohmian mechanics, has in fact been performed~\cite{nino}.\footnote{Since that analysis does not use the exact form of the law of motion, generalization seems to be easily attainable.}

According to~\cite{nino}, the spread of the particle wave function right after the position measurement\footnote{The conditional wave function for a certain configuration of the measurement device.} cannot be greater than the inaccuracy in the measurement. In fact, we may know the position of the particle only as accurately as the $\vert\psi_{\text{after}}\vert^2$-distribution allows, where $\psi_{\text{after}}$ is the wave function of the particle after the measurement. This result may be obtained either by analysing the generic process of the position measurement~\cite{nino} or by statistical arguments (see, e.g., Ch. 11 of~\cite{red_book}).\footnote{In Bohmian mechanics, this fact is referred to as \textit{absolute uncertainty}.} Therefore, $\epsilon_q\sim \Delta q$ does indeed hold. More precisely, if we determined the position of the particle to be $q_0$ and defined the accuracy of our measurement to be the number $\epsilon_q$ such that the probability of finding the particle in the region $B_{\epsilon_q}(q_0)$ is $1-\alpha$, then $\int_{B_{\epsilon_q}(q_0)}\vert\psi_{\text{after}}(q)\vert^2\mathrm{d}q=1-\alpha$.\footnote{We emphasize that this is not just the statement of the Born rule! The wave function after the measurement is determined by the wave function before the measurement and the interaction Hamiltonian. The accuracy of our measurement is determined by the interaction between the particle and the measurement device as well, but it is not obvious that we cannot infer the position of the particle from the output of the measurement device (`position of the pointer') more accurately than from $\psi_{\text{after}}$. That would be necessarily the case only in a $\psi$-complete theory.}   

The smaller the spread of the wave function, the faster it broadens in time, and the faster the volume of the image $U_\tau$ of the region $B_{\epsilon_q}(q_0)$ grows under the Bohmian flow $\Phi_t$, as $\int_{B_{\epsilon_q}(q_0)}\vert\psi_{\text{after}}(q)\vert^2\mathrm{d}q=1-\alpha=\int_{\Phi_\tau(B_{\epsilon_q}(q_0))}\vert\psi(q,\tau)\vert^2\mathrm{d}q$ has to hold due to equivariance~\cite[p. 152]{equivariance}. So, even if we know the system--measurement device interaction Hamiltonian and can calculate the wave function of the system after the measurement, we will be able to predict the future position of the particle only with low accuracy, because $U_\tau$ is going to be much larger than $\epsilon_q$. Heuristically, one could say that the rate of spread of the wave function corresponds to $\Delta p$ -- the spread of its Fourier transform. Thus, although Heisenberg's analysis in his Chicago lectures was far from being correct, his intuition was right after all.

\section{TCQT or a $\psi$-complete theory: a matter of taste?}

So, at least for Bohmian mechanics, `verification' of the presence of the trajectories in the sense of Heisenberg is indeed impossible. For a physicist who defines a physical theory as a set of rules allowing us to obtain numerical predictions for experiments, this is a convincing argument against TCQTs and, in particular, Bohmian mechanics. In classical physics the accurate prediction of the future position of a particle using its trajectory is indeed possible, so the idea of using the distribution of particles in space instead does seem very unnatural. Apart from that, the corresponding formalism (Newtonian/Lagrangian/Hamiltonian) is simpler than the one based on the density in phase space and Liouville's equation. In quantum physics, the accurate prediction of positions with trajectories is excluded and calculation of the trajectories (at least in Bohmian mechanics) requires the solution of the Schr\"odinger equation anyway, so for a physicist with an instrumentalist attitude, keeping only the wave function in the theory is preferable.

This is correct unless one takes into account that there is at least one big class of experiments for which standard quantum formalism, based on the wave function and self-adjoint operators, does not give unambiguous predictions: arrival time measurements. The general scheme of such measurements can be described as follows. A particle is first trapped in a certain region of space and then released at a known time, which is set to zero. A detector of given geometry is placed at a certain distance from the region of initial confinement. At time $\tau>0$, it clicks. This experiment is repeated many times and the distribution of the arrival times is acquired.    

The standard quantum formalism, it turns out, does not furnish definite predictions for the described experiments (see, e.g., \cite{muga}). Indeed, in 1933 Pauli~\cite{pauli} already noticed that there is no canonical time-operator in QM. As a result, over the years, many add-ons to the standard formalism aiming to predict the arrival time distributions of a quantum particle have been suggested~\cite{muga}. Not only is the adequacy of many of these proposals questionable, but also their range of applicability is severely limited~\cite{mielnik,leav,mugareply,leavreply,markus,ward}. Arguably, a generally applicable and internally-consistent recipe for describing quantum arrival-time experiments based solely on a $\psi$-complete theory is yet to be discovered. 
%To our best knowledge none of the proposed add-ons can provide a prediction for the whole variaty of the possible ToA experiments: with or without EM-field, one or several particles, presence of the potential barriers on the particle's way, different confinements and geometries of the detectors.

On the other hand, in any TCQT, prediction of the arrival time distribution is a problem with an almost obvious solution: if particles follow trajectories, the time when a given trajectory crosses a certain surface can be easily calculated. The arrival time distribution can thus be obtained if the distribution of the initial positions is known (see, e.g.,~\cite{siddhant}).

Thus even a physicist for whom a theory is solely a tool for obtaining numerical predictions for laboratory experiments may be willing to prefer a TCQT over a $\psi$-complete theory.

\section{Conclusion}

Let us summarize our discussion. Heisenberg's argument against particle trajectories in 1927 was not based on the uncertainty relation, but on his operationalist redefinition of the familiar physical concepts. This argument has been criticized as circular.

In contrast to what is often stated, Heisenberg's argument against trajectories was not related to the fact that the position and velocity cannot be defined simultaneously for a wave packet.

In his 1929 Chicago lectures, Heisenberg claimed that the existence of trajectories in the quantum world is impossible to `verify experimentally'. By that he meant that, because of his uncertainty relation~\eqref{kennard}, no matter how accurately we know the initial position of the particle, we will not be able to predict the future position with comparable accuracy. This is why, he claimed, to accept or to reject the trajectories is a matter of personal belief.

This argument may be appealing, but Heisenberg's derivation of the impossibility of the `experimental verification' 
%from the well-known uncertainty relation for standard deviations~\eqref{kennard} 
is irrelevant, since the TCQT he used was not empirically adequate. Nevertheless the analysis in Bohmian mechanics confirms that the `verification' in Heisenberg's sense is indeed impossible, at least in this trajectory-containing quantum theory. 

If one defines the physical theory as a set of rules to obtain predictions for experiments, this could be reason enough to discard Bohmian mechanics and opt for a $\psi$-complete quantum formalism, unless one takes into account the difficulties that $\psi$-complete theories encounter predicting the results of arrival time measurements. On the other hand, if one would like a physical theory to tell us about what there is, what the world consists of, how its constituents behave, and how this results in our observations, TCQTs and, in particular, Bohmian mechanics are definitely advantageous. 

\bmhead{Acknowledgments}

The author wishes to thank Dr. Paula Reichert, Siddhant Das, and Stephen N. Lyle for their help in the preparation of this paper.

%\bibliography{sn-bibliography}% common bib file
%% if required, the content of .bbl file can be included here once bbl is generated
%%\input sn-article.bbl

%% Default %%
%\input bib1.tex%

\end{document}